\documentclass[a4paper]{jpconf}

\usepackage{epsfig,graphicx,graphics,rotating}
\usepackage{dcolumn}
\usepackage{bm}
\usepackage{hyperref}
\usepackage{iopams}
\usepackage{setstack}
\usepackage{amsthm,amsfonts,amssymb}
\usepackage{upgreek}
\usepackage[latin1]{inputenc}
\usepackage{latexsym}

\begin{document}

\title{Testing effective quantum gravity with gravitational waves from extreme mass ratio inspirals}

\author{N.~Yunes$^{1}$ and C.~F.~Sopuerta$^{2}$}
\address{$^{1}$Institute for Gravitational Physics and Geometry,
         Center for Gravitational Wave Physics,
         Department of Physics, The Pennsylvania State University,
         University Park, PA 16802-6300}
\address{$^{2}$Institut de Ci\`encies de l'Espai (CSIC-IEEC), 
	Facultat de Ci\`encies, Campus UAB, Torre C5 parells, 
	Bellaterra, 08193 Barcelona, Spain.}
\ead{$^{1}$nyunes@princeton.edu}

%
\newcommand\be{\begin{equation}}
\newcommand\ba{\begin{eqnarray}}
\newcommand\ee{\end{equation}}
\newcommand\ea{\end{eqnarray}}

\newcommand{\mb}[1]{\mbox{\boldmath $#1$}}
\newcommand{\nn}{\nonumber}

\newcommand{\sch}{{\mbox{\tiny Sch}}}
\newcommand{\GW}{{\mbox{\tiny GW}}}
\newcommand{\GR}{{\mbox{\tiny GR}}}
\newcommand{\ZM}{{\mbox{\tiny ZM}}}
\newcommand{\RW}{{\mbox{\tiny RW}}}
\newcommand{\TT}{{\mbox{\tiny TT}}}
\newcommand{\MAT}{{\mbox{\tiny mat}}}
\newcommand{\STF}{{\mbox{\tiny STF}}}
\newcommand{\CPM}{{\mbox{\tiny CPM}}}
\newcommand{\CS}{{\mbox{\tiny CS}}}
\newcommand{\AXIAL}{{\mbox{\tiny Axial}}}
\newcommand{\POLAR}{{\mbox{\tiny Polar}}}
\newcommand{\MIN}{{\mbox{\tiny min}}}
\newcommand{\INC}{{\mbox{\tiny inc}}}
\newcommand{\APO}{{\mbox{\tiny apo}}}
\newcommand{\PERI}{{\mbox{\tiny peri}}}
\newcommand{\pont}{{\,^\ast\!}R\,R}
\newcommand{\met}{\mbox{g}}
\newcommand{\metb}{\mbox{\bf g}}
\newcommand{\oone}{{}^{\mbox{\tiny $(1)$}}}
\newcommand{\otwo}{{}^{\mbox{\tiny $(2)$}}}
\newcommand{\lgw}{\lambda^{}_{\GW}}
\newcommand{\hgw}{h^{}_{\GW}}

\begin{abstract}
  
Testing deviation of GR is one of the main goals of the proposed {\emph{Laser Interferometer Space Antenna}}.
For the first time, we consistently compute the generation of gravitational waves from extreme-mass ratio inspirals (stellar compact objects 
into supermassive black holes) in a well-motivated alternative theory of gravity, that to date remains weakly constrained by 
double binary pulsar observations. 
The theory we concentrate on is Chern-Simons (CS) modified gravity, a 4-D, effective theory that is motivated both 
from string theory and loop-quantum gravity, and which enhances the Einstein-Hilbert action through the addition of a dynamical 
scalar field and the parity-violating Pontryagin density. 
We show that although point particles continue to follow geodesics in the modified theory, the background about which they inspiral
is a modification to the Kerr metric, which imprints a CS correction on the gravitational waves emitted.  
CS modified gravitational waves are sufficiently different from the General Relativistic expectation that they lead to significant
dephasing after 3 weeks of evolution, but such dephasing will probably not prevent detection of these signals, but instead lead to a 
systematic error in the determination of parameters.  
We end with a study of radiation-reaction in the modified theory and show that, to leading-order, energy-momentum emission
is not CS modified, except possibly for the subdominant effect of scalar-field emission.
The inclusion of radiation-reaction will allow for tests of CS modified gravity with space-borne detectors that might be two
orders of magnitude larger than current binary pulsar bounds. 
  
\end{abstract}






\section{Motivation}
\label{intro}

One of the primary goals of the {\emph{Laser Interferometer Space Antenna}} (LISA) is to study possible
deviations from General Relativity (GR) in the strong field, where gravity is strong, speeds are large and binary
pulsar or Solar System experiments are inapplicable. One route to such tests has been to study deviations from 
the Kerr metric {\emph{within GR}}, through the determination of the multipolar structure of the gravitational field. 
In GR, however, the Kerr metric is known to be the only physically-reasonable representation of
the exterior gravitational field of spinning compact objects, provided cosmic censorship and causality holds. 

Another route to test GR is to consider alternative theories of gravity and study the imprint these leave on 
gravitational waves (GWs) emitted by LISA sources. Here we report on the first consistent study of GW generation
in a well-motivated alternative theory of gravity by the inspiral of a small compact object into a spinning, supermassive BHs 
(a so-called extreme-mass ratio inspiral or EMRI)~\cite{Sopuerta:2009iy}. 

Due to the plethora of alternative theories available, it becomes difficult to justify the choice of one theory over another. 
For this purpose, it is convenient to propose certain criteria for a theory
to be {\emph{a reasonable candidate}} to test GR with LISA~\cite{bias}: 
\begin{itemize}
\item {\emph{Metric theories of gravity}}: the gravitational field is described by
a symmetric metric tensor that satisfies the weak-equivalence principle~\cite{Misner:1973cw}. 
\item {\emph{Weak-field consistency}}: reduction to GR for weak gravitational fields
and small velocities, such that experimental and observational tests are passed.   
\item {\emph{Strong-field inconsistency}}: deviations from GR in the dynamical strong-field, 
where gravity is strong and speeds are large.
\end{itemize}
The weak-field consistency criterion also implies the existence and stability of physical solutions, such as
the Newtonian limit of the Schwarzschild metric to describe physics in the Solar System.
Additional criteria can also be proposed, such as well-posedness of the initial-value formulation, the existence
of a well-defined and complete relativistic action, and some type of motivation from fundamental physics. 
These additional criteria, however, might be too stringent since only GR has been sufficiently studied to determine
whether they are satisfied.

A modification to GR that passes all the criteria in the itemized list above is Chern-Simons (CS) modified
gravity~\cite{Alexander:2009tp}. In this 4D theory, the Einstein-Hilbert action is modified through the addition of the product
of a dynamical scalar field and the Pontryagin density. Since the scalar field couples only to the quadratic curvature correction
to the action, the weak-equivalence principle remains satisfied and the theory is indeed metric. Due to its quadratic curvature nature, 
this theory is also weak-field consistent but strong-field inconsistent. Stable solutions, like the Schwarzschild metric and a 
modified Kerr metric~\cite{Yunes:2009hc}, have also been found, although there are only qualitative arguments that suggest Cauchy 
well-posedness~\cite{Grumiller:2008ie}. As for motivation, the CS modification arises generically and unavoidingly in the 
low-energy effective limit of string theory, both in the perturbative sector due to the Green-Schwarz anomaly-canceling mechanism~\cite{Polchinski:1998rr}, and in the non-perturbative sector due to D-instanton charges and duality symmetries~\cite{Alexander:2004xd}. 
The CS correction can also be shown to arise in loop quantum gravity, when the Barbero-Immirzi parameter 
is promoted to a field in the presence of fermions~\cite{Taveras:2008yf,Mercuri:2009zt}. 
In fact, one can show generically (eg.~in cosmological inflation) that the CS correction is one of a few terms that generically 
arise in effective field theories when one expands the action in the curvature tensor~\cite{Weinberg:2008hq}.  

Two different version of CS modified gravity exist, a non-dynamical and a dynamical one, where the former has been strongly
constrained by theoretical and experimental considerations, while the latter has only bee weakly constrained by binary pulsar observations.
The non-dynamical version assumes the scalar field is {\emph{a priori}} prescribed and non-evolving, which forces what would have been an 
evolution equation for the scalar field into an additional constraint on the solution space of the theory. This additional constraint has been shown 
to be too restrictive, essentially disallowing spinning black hole (BH) solutions~\cite{Grumiller:2007rv} and forbidding perturbations of non-spinning 
BHs~\cite{Yunes:2007ss}. From an observational standpoint, the non-dynamical version has been strongly constrained in the neighbourhood
of neutron stars through binary pulsar observations~\cite{Yunes:2008ua}; GWs would allow a constrain in regions
of spacetime between source and observer due to corrections to GW propagation~\cite{Alexander:2007kv,Yunes:2008bu}.

Dynamical CS gravity has been only weakly constrained by binary pulsar observations~\cite{Yunes:2009hc}. Recently,~\cite{Cardoso:2009pk} studied 
perturbations of non-spinning, Schwarzschild BHs in the dynamical theory and found certain instabilities, which they used to argue a strong 
constraint could be placed due to the observationally confirmed existence of rotating BHs. 
This result, however, is strongly dependent on the assumption that there is no background scalar field to which
the metric perturbations could couple to. Such an assumption is not valid for rotating BHs, even in the limit of small rotation, as found
in~\cite{Yunes:2009hc}.  Then, the work of~\cite{Cardoso:2009pk} can be considered as a good motivation to explore in more
detail oscillations of BHs in the dynamical theory. As a first step, one would study the behaviour of {\emph{linear}} 
oscillations for rotating BHs (which has not been done yet as it can be prohibitively difficult).  Then, in the case 
that instabilities are still present, one could ask whether {\emph{non-linear}} effects can suppress the linear instabilities.  If the 
answer is in the negative, one can then try to put  constraints on the theory based on astrophysical observations.

We here study EMRIs in dynamical CS modified gravity. 
We first show that point-particles follow geodesics in this theory as in GR, although the supermassive BH background is 
modified.
Such a modification to geodesic motion leads to corrections in the generation of GWs that lead to a dephasing between CS waves and 
those expected in GR.
This dephasing will not prevent detection of GWs with LISA, but instead it will bias the estimation of parameters, leading to an uncontrolled systematic
error.
We conclude with a study of radiation-reaction in dynamical CS modified gravity and we proof that to leading order this is the same as in GR,
except perhaps for subdominant energy-momentum emission by the scalar field. 
The inclusion of radiation-reaction will lead to stronger GW modifications that will break the degeneracy between the CS coupling parameter
and the system parameters, thus allowing for a more stringent test of the dynamical theory relative to the binary pulsar
constraint.

The remainder of this Proceedings describes the above results in more detail:
Section~\ref{CS-basics} presents the basics of the theory; 
Section~\ref{Semi-relCS} describes the generation of GWs;
Section~\ref{Waves-Rad-Reac} shows the difference between GR and CS GWs and discusses the inclusion of radiation-reaction;
Section~\ref{Conclusions} concludes and points to future work.
We employ the conventions in Misner, Thorne and Wheeler~\cite{Misner:1973cw}, described in more 
detail in~\cite{Sopuerta:2009iy}. 

\section{Chern-Simons Basics}
\label{CS-basics}

We here summarize the basics of CS modified gravity and we refer the reader to the review paper~\cite{Alexander:2009tp}
for further details.

\subsection{Field Equations}

Consider then the Lagrangian density 
\be
\label{actions}
{\cal{L}}^{}_{} = \kappa R + \frac{\alpha}{4} \vartheta \; \pont - \frac{\beta}{2}  \left[ \met^{\mu \nu}
\left(\nabla_{\mu} \vartheta\right) \left(\nabla_{\nu} \vartheta\right) + 2 V(\vartheta) \right] + {\cal{L}}^{}_{\rm mat},
\ee
where the gravitational constant is $\kappa = (16 \pi G)^{-1}$; the coupling constant associated
with the gravitational CS correction is $\alpha$, while that associated with the strength of
the CS scalar field $\vartheta$ is $\beta$. The parity-violating Pontryagin density
$\pont :=  \epsilon^{\alpha \beta \mu \nu} R^{}_{\alpha \beta \gamma \delta} R^{\gamma \delta}{}^{}_{\mu \nu}/2$,
where the asterisk denotes the dual tensor, constructed using the antisymmetric Levi-Civita tensor 
$\epsilon^{\alpha\beta\mu\nu}$.  The potential term $V(\vartheta)$ will be assumed to vanish, 
as it does in string theory due to shift symmetry. The last term in~(\ref{actions}) represents the
Lagrangian density for external matter degrees of freedom.

Variation of the action with respect to all degrees of freedom leads to the CS field equations:
\ba
 G^{}_{\mu \nu} + \frac{\alpha}{\kappa} C^{}_{\mu \nu}  = \frac{1}{2 \kappa} \left(T_{\mu \nu}^{\MAT} 
+ T_{\mu \nu}^{(\vartheta)}\right)\,,  \label{EEs} \\
 \beta \square \vartheta = \beta \frac{d V}{d\vartheta} - \frac{\alpha}{4} \pont\,,
\label{EOM}
\ea
where $T_{\mu \nu}^{\MAT}$ is a matter stress-energy tensor, $T_{\mu \nu}^{(\vartheta)}$ is the stress-energy of the 
CS scalar field and $C^{\mu \nu}$ is the so-called C-tensor, both of which are given by 
\ba
T_{\mu \nu}^{(\vartheta)} = \beta \left[ (\nabla_{\mu} \vartheta) (\nabla_{\nu}\vartheta) - \frac{1}{2} 
\met_{\mu \nu}(\nabla^{\sigma} \vartheta) (\nabla_{\sigma} \vartheta) -  \met_{\mu \nu}V(\vartheta) \right]\,.
\label{vartheta-Tab}
\nonumber \\
C^{\alpha \beta} = \left(\nabla^{}_{\sigma} \vartheta\right) \epsilon^{\sigma \delta \nu(\alpha}
\nabla^{}_{\nu}R^{\beta)}{}^{}_{\delta} + 
\left(\nabla^{}_{\sigma}\nabla^{}_{\delta}\vartheta \right) 
{\,^\ast\!}R^{\delta (\alpha \beta)\sigma}\,.
\label{Ctensor}
\ea

As we are here interested in the dynamical version of dynamical CS modified gravity, we restrict attention to coupling
constants $\beta \neq 0$. If $\beta = 0$, one would recover non-dynamical CS theory and 
the equation of motion for $\vartheta$ would reduce to the Pontryagin constraint $\pont = 0$, 
which over-constrains the field equations~\cite{Yunes:2007ss}. We
consider the dynamical theory, where we define $\xi := \alpha^{2}/(\kappa \beta)$ as an expansion parameter 
for the so-called small-coupling approximation (see~\cite{Yunes:2009hc} for details), in which a modified Kerr
solution is obtained.

\subsection{Gravitational Waves in CS theory}

Let us decompose the metric tensor as 
$\met^{}_{\mu\nu} = \bar{\met}^{}_{\mu\nu} + \epsilon \;  h^{}_{\mu\nu}\ + {\cal{O}}(\epsilon^{2})$,
where $\bar{\met}^{}_{\mu \nu}$ is some background metric, $h^{}_{\mu \nu}$ is a metric perturbation, 
which in the radiation-zone (many GW wavelengths away from the center of mass) can be associated with GWs, 
and $\epsilon$ is a book-keeping perturbation parameter associated with the GW strength.

The CS field equations can now be expanded in $\epsilon$. In the radiation zone, however, we can further
simplify the analysis by focusing on a flat background, $\bar{\met}^{}_{\mu \nu} =  \eta_{\mu \nu}$. We then
find 
\ba
 \label{lin-mod-FE}
 \fl
- \frac{1}{\kappa} \bar{T}_{\mu \nu} =
\epsilon \square_{\eta} h_{\mu \nu} 
+ \frac{\alpha}{\kappa} (\partial_{\sigma} \vartheta) \epsilon \; \bar\epsilon^{\,\sigma \delta \xi}{}_{(\mu} 
\square_{\eta} h_{\nu) \delta,\xi}  
\nonumber \\
+ \frac{\alpha}{\kappa} (\partial_{\sigma}{}^{\gamma} \vartheta) \epsilon \; \bar\epsilon^{\,\sigma\delta \xi}{}_{(\mu} 
\left[ h_{\xi \gamma,|\nu) \delta} - h_{\nu) \xi,\gamma \delta} \right]\,
+ {\cal{O}}(\epsilon^{2}) 
\\
\fl
 \beta \square \vartheta = -\frac{\alpha}{2} \epsilon^{2} \;  \bar{\epsilon}^{\,\alpha \beta \mu \nu} h_{\alpha \delta,\gamma \beta} 
h_{\nu}{}^{[\gamma,\delta]}{}_{\mu} + {\cal{O}}(\epsilon^{3}) \,, \label{CS-FE}
\ea
where $\bar{T}_{\mu \nu} = T_{\mu \nu} - (1/2) \eta^{}_{\mu \nu} T$ is the 
trace-reversed version of the total stress-energy tensor 
$T_{\mu \nu} = T_{\mu \nu}^{\rm mat} + T_{\mu \nu}^{(\vartheta)}$, 
$\square_{\eta} = \eta^{\mu \nu} \partial_{\mu} \partial_{\nu}$ is the flat-space 
D'Alembertian and $\bar{\epsilon}^{\,\mu \nu \alpha \beta}$ is the flat-space Levi-Civita tensor density
(not to be confused with $\epsilon$). We work in Lorenz gauge,  
$\left(h^{\mu\nu} - \frac{1}{2}h \bar{\met}^{\mu\nu}\right)^{}_{|\nu}  = 0$, 
where $h = \bar{\met}^{\mu \nu} h_{\mu \nu}$ is the trace of $h^{}_{\mu \nu}$ with respect to $\bar{\met}^{}_{\mu \nu}$, 
and we restrict the gauge further to a transverse-traceless one by $h = 0$.

The GW generation formula are then identical to the GR ones to leading order in $\epsilon$, as one can see
in~(\ref{lin-mod-FE}). This is because the CS correction is proportional both to the scalar field and the metric perturbation, 
where the former is sourced by the Pontryagin density [see~(\ref{CS-FE})], which itself is of ${\cal{O}}(\epsilon^{2})$ in the 
radiation zone. One can make this argument more formally accurate by decomposing the scalar field itself into a background
plus a perturbation and then performing a multiple-scale perturbative expansion, but we relegate such details to~\cite{Sopuerta:2009iy}.
Furthermore, one can show that there are only two independent GW polarizations in dynamical CS gravity, 
as is the case in GR, in spite of the extra scalar field present in the theory. This can be done by studying the geodesic
deviation equation or plane wave solutions, as found in~\cite{Sopuerta:2009iy,Grumiller:2008ie}.

\section{EMRIs in CS Gravity}
\label{Semi-relCS}

We now consider the generation of GWs by EMRIs in dynamical CS gravity. We shall approximate
these waves in the so-called semi-relativistic approximation~\cite{Ruffini:1981rs}, where the motion is assumed geodesic 
and GWs are assumed to propagate in flat spacetime.
This simplifying assumptions have recently been shown to be accurate
relative to more precise black-hole perturbation theory calculations~\cite{Babak:2006uv}.
We begin with a description of the background upon which geodesics are evolved. We then show that point-particle orbits
can truly be described with geodesics and explicitly derive the latter. We end by presenting the GW generation formula
in the semi-relativistic approximation.

\subsection{Background Geometry}

Via the small-coupling approximation, complimented by a slow-rotation approximation, one can solve for the exterior
gravitational field of a rotating BH in dynamical CS modified gravity. The line element is given by 
$ds^{2} = ds^{2}_{\rm Kerr} + 5 \xi a/(4 r^{4}) \left[1 + 12 M/(7 r) + 27 M^{2}/(10 r^{2}) \right] \sin^{2}{\theta} dt d\phi$, 
where $ds^{2}_{\rm Kerr}$ is the line element for the Kerr metric and we have used Boyer-Lindquist type 
coordinates~\cite{Yunes:2009hc}. 

The CS modified metric remains stationary and axisymmetric, allowing for the definition of multipole moments appropriately. 
The multipolar structure of the modified metric remains completely
determined by only two moments (no-hair or two-hair theorem): the mass monopole and the current dipole. The relation, however, 
between these two moments and higher order ones is modified from the GR expectation at multipole $\ell > 4$. In spite of the CS scalar field, 
the dynamical theory satisfies the no-hair theorem, as the scalar field is
fully determined by the geometry.  

\subsection{Motion of Massive Bodies}

One can show that test-bodies follow geodesics in CS modified gravity~\cite{Sopuerta:2009iy}. This is 
because the divergence of the field equations~(\ref{EEs}) leads to~(\ref{EOM}) for the non-matter degrees
of freedom, forcing the divergence of the matter stress energy tensor to vanish. This then proves dynamical
CS modified gravity satisfies the weak-equivalence principle, thus rendering it a metric theory as we anticipated
in the Introduction.

The motion of small compact object with an extreme-mass ratio can then be modeled, to zeroth order, by 
geodesics in the modified Kerr background of the previous section. As the background remains stationary 
and axisymmetric, there still exist timelike and azimuthal, commuting Killing vectors that define a conserved
energy $E$ and angular momentum $L$ per unit mass. Moreover, the modified background also possesses
an additional constant of the motion (the Carter constant, $Q$) associated with a Killing tensor. Using these
constants and the orthonormality condition for timelike geodesics, we can derive the geodesic equations
$\dot{x}^{\mu} = \dot{x}^{\mu}_{\rm Kerr} + \delta x^{\mu}_{\CS}$, where an overhead dot stands for partial 
time derivative and $x^{\mu} = [t,r,\theta,\phi]$ are Boyer-Lindquist coordinates (where the polar angle $\theta$ 
is not to be confused with the CS scalar field $\vartheta$). The correction factor 
$\delta x^{\mu}_{\CS} = [L, 2 E L, 0, -E] \delta g^{\CS}_{\phi}$, where
$\delta g_{\phi}^{\CS} = \xi a/(112 r^{8} f) (70 r^{2} + 120 r\,M + 189 M^{2})$ and $f := 1 - 2 M/r$. 

The geodesic equations allow one to compute some physical observables that waveforms are sensitive to. 
One such observable is the innermost-stable circular orbit (ISCO) location, which is CS shifted by~\cite{Yunes:2009hc}:
$R^{}_{\mbox{\tiny ISCO}}= 6M \mp 4\sqrt{6}a/3 - 7a^2/(18 M) \pm 77\sqrt{6}a \xi/(5184 M^{4})$,
where the upper (lower) signs correspond to co-rotating (counter-rotating) geodesics. Notice that the 
CS correction works {\emph{against}} the spin effects.
Other observables of interest are the fundamental orbital frequencies associated with orbital motion~\cite{Schmidt:2002qk,Drasco:2003ky}. 
In CS theory, the temporal, radial and azimuthal frequencies are modified in an unilluminating way that we avoid 
here~\cite{Sopuerta:2009iy}, if one employs the time coordinate introduced in~\cite{Mino:2003yg}. 
Therefore, CS geodesics correspond to GR geodesics with different fundamental frequencies, which maps to different
system parameters. 
 
\subsection{GWs for EMRIs}

The first simplification of the semi-relativistic approximation (that of modeling orbital motion as geodesic) has
already been discussed in the previous section, and so we here concentrate on the second semi-relativistic simplification:
GWs propagating on flat spacetime. We implement this approximation by modeling the solution of the linearized GW generation 
formula in~\ref{lin-mod-FE} via a multipolar expansion (see eg.~\cite{Thorne:1980rm}). In particulat, we employ a 
quadrupole-octopole approximation, where the plus- and cross-polarizations of the waveforms are given by
\be
h^{}_{+,\times} = \frac{1}{r}\varepsilon^{ij}_{+,\times} \left[ \ddot{I}_{ij} - 2 n^{\mu} \ddot{S}_{\mu i j} + n^{\mu} \dddot{I}_{\mu i j} \right], 
\ee
where $r$ is the flat-space distance from the source to the observer, $n^{\mu} = [1,x^{i}/r]$, $\epsilon^{ij}_{+,\times}$ are polarization tensors, 
and $I_{ij}$, $S_{ijk}$ and $I_{ijk}$ are the symmetric and trace-free projections of the mass quadrupole, current octopole and 
mass octopole respectively, defined as integrals of the product of point-particle stress-energy tensor with the point-particle trajectories
(see eg.~\cite{Yunes:2008gb} for explicit expressions). 

An important detail in the above implementation is the relation between the coordinates used in the geodesic evolution (Boyer-Lindquist)
and those employed in the metric reconstruction (Cartesian). We here choose to identify Boyer-Lindquist coordinates 
with flat-space spherical coordinates, such that $(x,y,z)=r (\sin\theta\cos\phi,\sin\theta\sin\phi,\cos\theta)$. Another important detail is
that the above approximation is truly a slow-motion one, which implies waveforms are accurate provided the pericenter orbital
velocity is much smaller than the speed of light. In spite of this, the semi-relativistic approximation does capture 
the correct {\emph{qualitative}} behaviour of EMRI orbits and waveforms.

\section{Testing CS theory with LISA and Radiation-Reaction}
\label{Waves-Rad-Reac}

We here describe some of the differences between a GR and a CS EMRI waveform. 
For this, we shall concentrate on the following test orbits:
(A) $(a,\xi,r^{}_{\PERI},e,\theta^{}_{\INC}) = (0.1\,M, 0.1 M^{4}, 12\,M, 0.2, 0.1)\,$, 
(B) $(a,\xi,r^{}_{\PERI},e,\theta^{}_{\INC}) = (0.2\,M, 0.2 M^{4}, 8\,M, 0.4, 0.2)\,$, 
(C) $(a,\xi,r^{}_{\PERI},e,\theta^{}_{\INC}) = (0.4\,M, 0.4 M^{4}, 6\,M, 0.6, 0.3)\,$,  
where $r^{}_{\PERI}$ stands for pericenter distance, $e$ for orbital eccentricity
and $\theta^{}_{\INC}$ is related to the inclination angle. Waveforms are measured by $z$-axis observers 
at a distance of $8\,$kpc and test orbits have been ordered from least to most relativistic. 
All orbits assume the central BH has $M=M^{}_{\bullet}$, 
where $M^{}_{\bullet} = 4.5 \times 10^{6} M_{\odot}$ is approximately the mass 
of the BH at the center of the Milky Way, while the SCO has $m=35 M_{\odot}$,
which leads to a mass ratio $\mu \sim 7.8 \times 10^{-6}$. We conclude with a discussion of 
radiation-reaction in CS theory.

\subsection{Trajectory and Waveform Dephasing}

The trajectories are obtained by solving the geodesic equations discussed above for a total time of $T = 5 \times 10^{5} M$, 
obtaining on the order of $10^{3}$-$10^{4}$ cycles.  (see eg.~\cite{Sopuerta:2009iy} for details). 
Geodesics are initialized with the same orbital data when performing GR and CS evolutions. That is, given some
set of conserved quantities $(E,L,Q)$, one can derive three initial orbital parameters $(p,e,\theta^{}_{\INC})$, but these will differ in
GR and CS due to the latter's geodesic corrections. We here consider orbits 
with the {\emph{same}} orbital parameters, instead of the same constants of motion.

\begin{figure}[htp]
\begin{center}
\includegraphics[scale=0.31,clip=true]{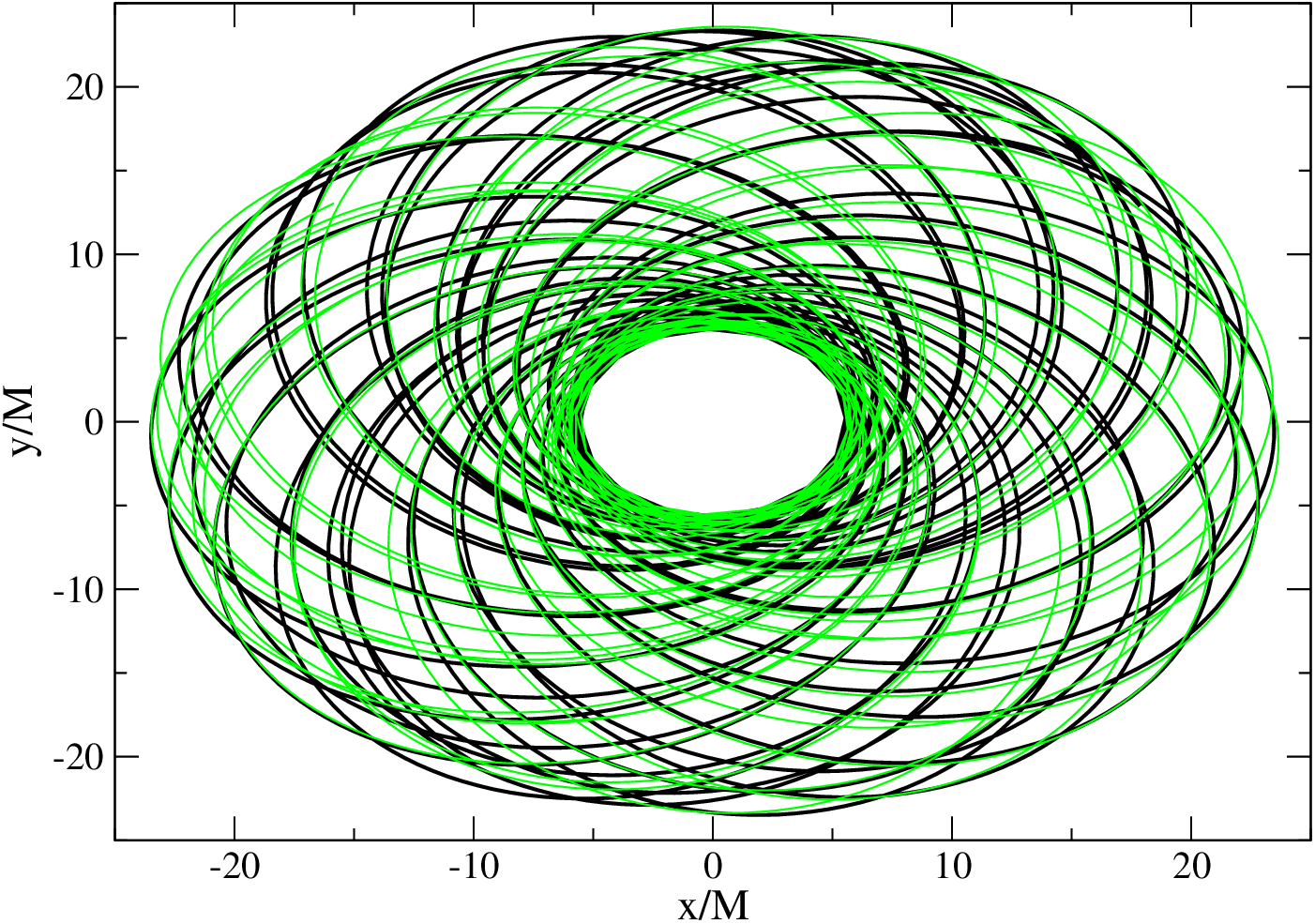}
\includegraphics[scale=0.31,clip=true]{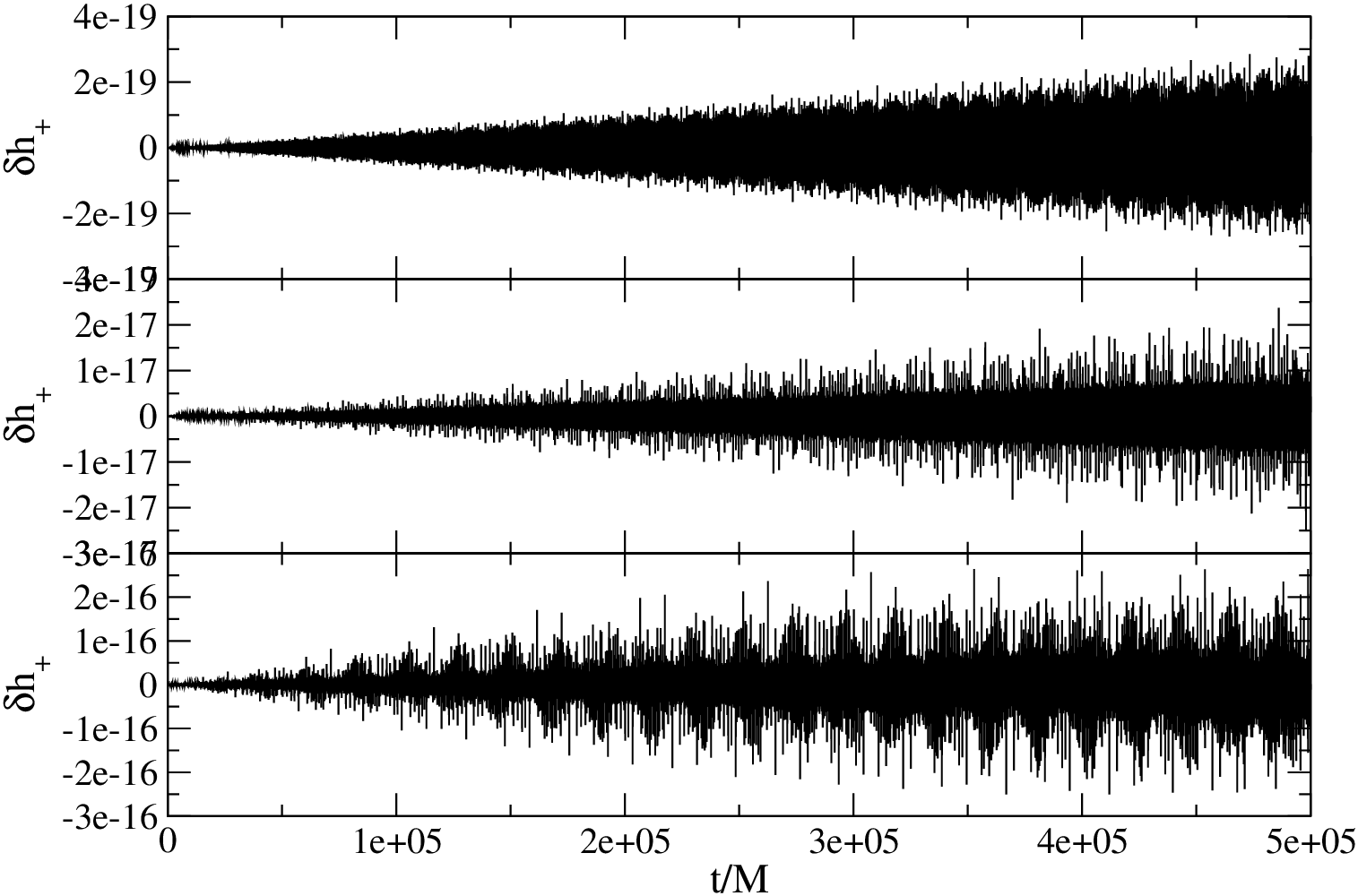}
\caption{\label{xy-plane} 
Left: Two-dimensional projection onto the $x$-$y$ plane of orbit $C$
in a Kerr background (light gray) and in the CS corrected Kerr metric (black). Observe that a large
dephasing has built up by the end of the evolution.
Right: Plus-polarized GW dephasing as a function of time. 
The top panel shows results for orbit $A$, the middle one for $B$ and the bottom one for $C$.
Observe that after 2/5 of the evolution, the waveform has dephased significantly.}
\end{center}
\end{figure}

The orbits chosen present rather generic behaviour with a stage of zoom-whirl, where the particle whirls violently for several cycles
close to pericenter and then zooms out to large radius only to return again close to pericenter, during all of which there is both 
in-plane and out-of-the-plane precession.  
The left panel of~\ref{xy-plane} shows the projection of the orbital trajectory onto the $x$-$y$ plane (orbit C) assuming a CS modified
Kerr background BH (black line) and a Kerr background (light gray line). We present only the last $17500 M$ of geodesic 
evolution, during which we can clearly observe that the trajectories have dephased significantly .   
We find that orbit $C$ presents the biggest dephasing, followed by $B$ and 
then $A$.

The right-panel of~\ref{xy-plane} shows the difference in the plus-polarized waveforms in CS gravity and GR.
As a reference, the maximum magnitude of the GR GW polarization
$h^{}_{+}$ is approximately $|h^{}_{+}| < 8 \times 10^{-17}$ (orbit $A$), $|h^{}_{+}| < 1.25 
\times 10^{-16}$ (orbit $B$), and $|h^{}_{+}| <  2.25 \times 10^{-16}$ (orbit $C$).
Observe that by the end of the simulation ($\sim 128$ days of data using $M = M^{}_{\bullet}$), orbit 
$A$ waveforms have dephased by $0.3 \%\,$, while orbit $B$ waveforms have dephased by $16 \%$, 
and orbit $C$ waveforms by $90 \%$ relative to the maximum amplitude of the respective GR waveforms. 
Similar behaviour is observed for the other polarization.

\subsection{Radiation-Reaction in Dynamical CS Gravity}

Radiation-reaction can be incorporated via a radiative, adiabatic 
approximation~\cite{Mino:2003yg,Sago:2005gd,Hughes:2005qb}, 
where the constants of the motion are evolved using balance laws that relate orbital energy
loss to GW energy emission. Such an approximation is valid provided the change in the 
constant of the motion occurs on a much larger timescale relative to the orbital one. 

The balance laws can be obtained by studying the effective stress-energy tensor associated with GWs
in the framework of the short-wavelength approximation~\cite{Isaacson:1968zz,Isaacson:1968zza}. 
We thus decompose both the geometry and the CS scalar field into a slowly- or non-oscillating background 
and a rapidly-oscillating perturbation. The linear part of the modified field equations describe the evolution
of the perturbation, while the averaged second-order part describes how the background is modified
by GWs. This later part serves as an effective GW stress-energy tensor, $T^{\GW}_{\mu\nu}$ (the Isaacson tensor),
\ba
\fl
T^{\GW}_{\mu\nu}  =  -2\kappa\left\{ < \otwo R^{}_{\mu\nu}[h]>
-\frac{1}{2}\bar{\met}^{}_{\mu\nu}< \otwo R[h]> 
\right. \nn \\  +  \left. \!\!\! \frac{\alpha}{\kappa}\left( <\otwo C^{}_{\mu\nu}[\bar\vartheta\,,h]> + 
<\oone C^{}_{\mu\nu}[\tilde\vartheta\,,h]>\right) \right\}\,
\label{cstgw}
\ea
where the superscript preceding any given quantity stands for the
perturbative order of that quantity with respect to metric perturbations.
Restricting attention to the transverse-traceless gauge and using properties of the averaging procedure, 
one can show that all terms associated with the CS correction in~\ref{cstgw} identically vanish, yielding the 
same expressions for the Isaacson tensor as in GR. 

The above result proofs that orbital backreaction in dynamical CS modified gravity is {\emph{similar}} but not identical to that expected in GR. 
First, in addition to GW emission, orbital energy is also lost via scalar field emission, although
the latter is a subleading effect. 
Second, even if one neglects such scalar field emission, GW emission depends on time-derivatives of the waveforms, which we showed
in the previous subsection are different in CS gravity. 

\subsection{A possible LISA Implementation of a CS Test}
 
What is the effect of the CS correction in LISA data analysis? As shown through the semi-relativistic waveforms, 
GR and CS waveforms dephase given sufficient time, but such dephasing will not prevent detection. Instead
such dephasing will only affect parameter estimation. That is, the CS correction modifies the fundamental geodesic
frequencies, and thus, a CS geodesic is equivalent to a GR geodesic with different orbital parameters. 

\begin{figure}[htp]
\begin{center}
\includegraphics[scale=0.4,clip=true]{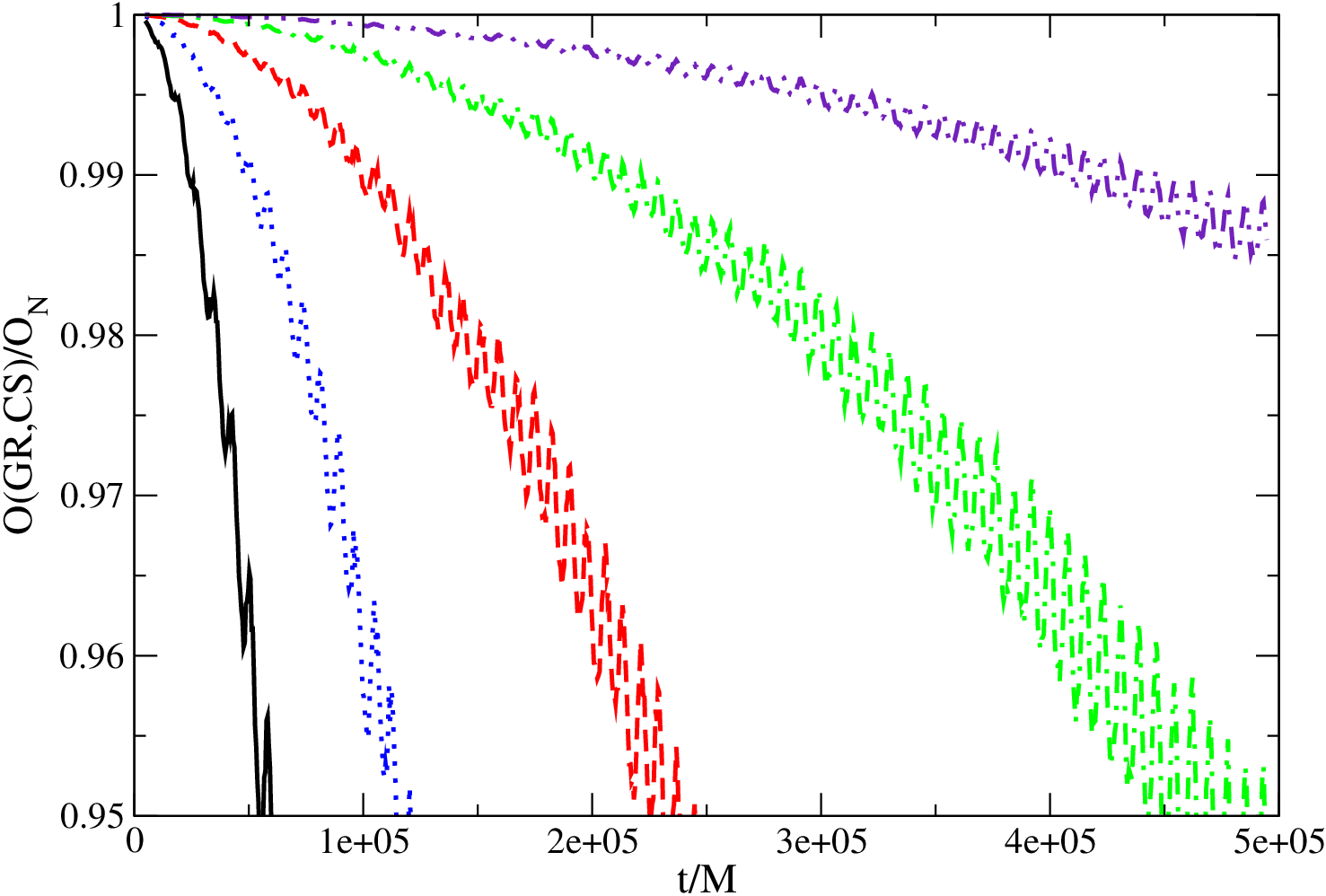}
\caption{\label{overl2} Normalized dephasing measure ${\cal{O}}(\GR,\CS)$ 
as a function of time for orbit $C$ with: 
$\xi = 0.4 M^{4}$ (solid black),
$\xi = 0.2 M^{4}$ (dotted blue),
$\xi = 0.1 M^{4}$ (dashed red), 
$\xi = 0.05 M^{4}$ (dot-dashed green),
and $\xi= 0.025 M^{4}$ (dot-dot-dashed violet). Observe that a significant dephasing is found even 
for $\xi = 10^{-2}$.}
\end{center}
\end{figure}

How big should the CS coupling parameter be to induce a sufficiently large dephasing that would 
contaminate parameter estimation?~\ref{overl2} shows the average of a normalized dephasing measure as a 
function of time for orbit $C$ with different $\xi$ (see caption).
%
%
This dephasing measure is nothing but the integrand of the overlap with averaged beam-pattern functions in the time-domain:
$\bigcirc(\GR,\CS) := [h^{\GR}_{+} h^{\CS}_{+} + h^{\GR}_{\times} h^{\CS}_{\times}]$ normalized by
$\bigcirc^{}_{N}(\GR,\CS) := \sqrt{\bigcirc(\GR,\GR)\,\bigcirc(\CS,\CS)}$. This figures shows that 
the CS correction could lead to a significant dephasing, and thus bias in parameter estimation, even for a CS parameter 
of ${\cal{O}}(10^{-2})$ after only $4$ months of data. 

The inclusion of radiation-reaction is crucial for breaking the degeneracy between the CS correction $\xi$ and the system
parameters, {\emph{i.e.}}~to distinguish between a GR and a CS geodesic. One expects that radiation-reaction will not
only break this degeneracy, but also lead to larger deviation from the GR expectation. A rough estimate of the accuracy
to which CS gravity could be constrained via a LISA observation is
$\xi^{1/4} \lesssim 10^{5}\, {\textrm{km}} \left(\delta/10^{-6}\right)^{1/4} \left(M/M_{\bullet}\right)$,
where $\delta$ is the accuracy to which $\xi$ can be measured, which depends on the integration time, 
the signal-to-noise ratio, the type of orbit considered and how much radiation-reaction affects the orbit. 
Notice that intermediate-mass ratio inspirals (IMRIs) are favored over EMRIs. This result is to be compared 
with the binary pulsar constrained $\xi^{1/4} \lesssim 10^{4} \; {\textrm{km}}$~\cite{Yunes:2009hc}. 
We then see that an IMRI with $M = 10^{3} M_{\odot}$ could place a constraint two-orders of magnitude more
stringent than the binary pulsar one. Moreover, a GW test can constrain the dynamical behaviour of the theory
in the neighbourhood of BHs, which is simply not possible with binary pulsar observations. 

\section{Conclusions}
\label{Conclusions}

Dynamical CS gravity is a viable theory that can only be constrained dynamically via GW observations. 
We have constructed geodesics in the modified theory and derived radiation-reaction formula that should
allow for the construction of CS-modified EMRI waveforms. Such waveforms could be used in LISA data analysis
pipelines to test the theory at least two-orders of magnitude better than current binary pulsar observations.

\ack
  
We thank Leor Barack and Jonathan Gair for comments and suggestions. 
We acknowledge support from NSF grant PHY-0745779,
the Ram\'on y Cajal Programme (MEC, Spain), by a Marie Curie
International Reintegration Grant (MIRG-CT-2007-205005/PHY) within the
7th European Community Framework Programme, and from the contract ESP2007-61712
of MEC, Spain.

\section*{References}
 
\bibliographystyle{iopart-num} 
\bibliography{master}

\end{document}